\documentclass[prl,twocolumn,showpacs,superscriptaddress,nofootinbib]{revtex4}
\usepackage{amsmath}
\usepackage[thickspace,cdot]{SIunits}
\usepackage[dvips]{hyperref}
\usepackage[dvips]{graphicx}
\usepackage{cancel}
\usepackage{ulem}
\usepackage{color}

\usepackage[latin1]{inputenc}
\usepackage{times}
\newcommand{\iA}{\ensuremath{\angstrom^{_{-1}}}}

\begin{document}
\preprint{submitted to Phys. Rev. Lett.}
\title{Linear plasmon dispersion in single-wall carbon
nanotubes\\ and the collective excitation spectrum of graphene}

\author{C. Kramberger}
\affiliation{IFW Dresden, Helmholtzstr. 20, D-01069 Dresden, Germany}

\author{R. Hambach}
\affiliation{Laboratoire des Solides Irradi{\'e}s, CEA-CNRS UMR
  7642-Ecole Polytechnique, 91128 Palaiseau Cedex, France and European
  Theoretical Spectroscopy Facility (ETSF), Palaiseau, France}
\affiliation{IFTO, Friedrich-Schiller-Universit{\"a}t Jena,
  07743 Jena, Germany}

\author{C. Giorgetti}
\affiliation{Laboratoire des Solides Irradi{\'e}s, CEA-CNRS UMR
  7642-Ecole Polytechnique, 91128 Palaiseau Cedex, France and European
  Theoretical Spectroscopy Facility (ETSF), Palaiseau, France}

\author{M. H. R\"ummeli}
\author{M. Knupfer}
\affiliation{IFW Dresden, Helmholtzstr. 20, D-01069 Dresden,
Germany}
\author{J. Fink}
\affiliation{IFW Dresden, Helmholtzstr. 20, D-01069 Dresden,
Germany}
\affiliation{BESSY I\hspace{-.1em}I,D-12489 Berlin,
Germany}

\author{B. B\"uchner}
\affiliation{IFW Dresden, Helmholtzstr. 20, D-01069 Dresden, Germany}

\author{L. Reining}
\affiliation{Laboratoire des Solides Irradi{\'e}s, CEA-CNRS UMR
  7642-Ecole Polytechnique, 91128 Palaiseau Cedex, France and European
  Theoretical Spectroscopy Facility (ETSF), Palaiseau, France}

\author{E. Einarsson}
\author{S. Maruyama}
\affiliation{The Univ. of Tokyo, Dept. of Mech. Eng., 7-3-1 Hongo,
Bunkyo-ku, Tokyo 113-8656, Japan}

\author{F. Sottile}
\affiliation{Laboratoire des Solides Irradi{\'e}s, CEA-CNRS UMR
  7642-Ecole Polytechnique, 91128 Palaiseau Cedex, France and European
  Theoretical Spectroscopy Facility (ETSF), Palaiseau, France}

\author{K. Hannewald}
\affiliation{IFTO, Friedrich-Schiller-Universit{\"a}t Jena,
  07743 Jena, Germany}

\author{V. Olevano}
\affiliation{Institut N\'eel, Grenoble, France and
  European Theoretical Spectroscopy Facility (ETSF), Palaiseau, France}

\author{A.G. Marinopoulos}
\affiliation{Laboratoire des Solides Irradi{\'e}s, CEA-CNRS UMR
  7642-Ecole Polytechnique, 91128 Palaiseau Cedex, France}
\affiliation{Department of Physics and Astronomy, Vanderbilt
  University, Nashville, TN 37235, USA}

\author{T. Pichler}
\affiliation{IFW Dresden, Helmholtzstr. 20, D-01069 Dresden,
Germany}
\affiliation{Institute of Materials Physics, University
of Vienna, Strudlhofgasse 4, A-1090, Vienna, Austria}

\date{submitted to Phys. Rev. Lett., \today}

\begin{abstract}
  We have measured a strictly linear $\pi$ plasmon dispersion along
  the axis of individualized single wall carbon nanotubes, which is
  completely different from plasmon dispersions of graphite or bundled
  single wall carbon nanotubes. Comparative \textit{ab initio} studies
  on graphene based systems allow us to reproduce the different
  dispersions. This suggests that individualized nanotubes provide
  viable experimental access to collective electronic excitations
  of graphene, and it validates the use of graphene to understand
  electronic excitations of carbon nanotubes. In particular, the
  calculations reveal that local field effects (LFE) cause a mixing of
  electronic transitions, including the \lq Dirac cone\rq, resulting
  in the observed linear dispersion.
\end{abstract}


\maketitle

Single-wall carbon nanotubes (SWNT) and its parent compound
graphene are archetypes of low dimensional systems with strongly
anisotropic and unique electronic properties which make them
interesting for both fundamental research and as building blocks
in nanoelectronic applications \cite{Avouris2006}. Their
electronic bandstructure is frequently studied. In graphene, the
linear band dispersion at the Fermi level, the \lq Dirac cone\rq,
leads to unique characteristics in nanoelectronic devices
\cite{Geim2007}. One can expect a strong analogy between graphene
and isolated SWNT for excitations along the sheet and along the
tube axis, respectively. Within the zone-folding model, i.e.
neglecting curvature effects, the graphene bandstructure is sliced
along parallel lines when the sheet is rolled up into a cylinder.
The result are characteristic van Hove singularities (VHS) in the
density of states (DOS)\cite{Hamada1992}. Bulk (i.e. bundled) SWNT
show an optical absorption peak at $\sim\unit{4.5}\electronvolt$
due to transitions of the $\pi$ electrons \cite{Kataura1999}. In
vertically aligned SWNT (VA-SWNT) one finds the same peak position
for on-axis polarization and an additional peak for perpendicular
polarization at $\sim\unit{5.2}\electronvolt$
\cite{Murakami2005a}. Further information can be obtained from
collective electronic excitations (plasmons) beyond the optical
limit \cite{Stephan2002} (i.e. momentum transfer $q > 0$). Angle
resolved electron energy loss spectroscopy (EELS) assesses the
detailed plasmon dispersion \cite{Pichler1998,Liu2001}, but it is
so far missing for freestanding isolated sp$^2$ carbon systems.
\par
Models based on the homogeneous electron gas \cite{Longe1993}, or
the tight-binding scheme \cite{Huang1997,Shyu2000} have been used
to describe these excitations. The former are however bound to
metallic systems. The latter have provided valuable insight and
predictions for the properties of isolated sheets, tubes, and
assemblies of these objects; in particular, they have predicted an
almost linear plasmon dispersion for isolated systems. However,
the tight binding results neglect screening beyond the $\pi$
bands, and they depend on parameters that hide the underlying
complexity. No realistic parameter-free calculations have been
performed to predict the plasmon dispersion in these systems, nor
has its origin been analyzed. Instead, \textit{ab initio}
spectroscopy calculations have dealt with absorption spectra
($q\!\to\! 0$) for SWNT
\cite{Marinopoulos2003,Spataru2004,Chang2004}, and plasmon
dispersions in graphite \cite{Marinopoulos2002,Marinopoulos2004};
most other available calculations are \textit{ground-state} or
\textit{bandstructure} ones \cite{Charlier2007}. The prediction,
comparison and interpretation of the full dispersive electronic
excitations of isolated SWNT and graphene sheets calls for new
experiments and for \textit{ab initio} theoretical support going
beyond bandstructure calculations.
\par
Indeed, electronic excitations imply a self-consistent response of
the entire system and have therefore to be described in terms of
bandstructure \textit{and} induced potentials. In solids, the
latter consist of microscopic induced components (local field
effects \cite{Baroni1987}) and a macroscopic induced component.
The latter is responsible for the difference between interband
transitions as measured in absorption, and plasmons as measured in
loss spectroscopies \cite{Sottile2005}. In isolated systems, the
macroscopic component will naturally vanish.  Absorption peaks and
loss peaks at $q\!\to\! 0$ will hence coincide. The microscopic
part, instead, can become dramatically important when macroscopic
screening is low.  Therefore, theory based on bandstructure alone
will be unable to describe isolated systems, whereas experiments
performed on bundles or graphite are not representative for
isolated SWNT and graphene. There is hence an important gap in our
understanding of the properties of these isolated sp$^2$ carbon
systems.
\par
The present work is meant to bridge this gap with a detailed EELS
study on freestanding mats of VA-SWNT, and corresponding {\it ab
  initio} calculations for graphene sheets. Our studies allow us to
give answers to several important questions, by \textit{ (i)
  distinguishing a localized perpendicular and a strictly linear
  on-axis plasmon dispersion in isolated single-wall nanotubes; (ii)
  showing the quantitative similarity to electronic excitations in
  graphene; (iii) analyzing the impact of local field effects on the
  linear plasmon dispersion, in particular the mixing-in of low-energy
  transitions; (iv) quantifying the importance of interactions
  between neighboring sheets or tubes.}
\par
$\unit{2}\micro\meter$ to $\unit{7}\micro\meter$ thick VA-SWNT
material was directly grown by catalytic decomposition of alcohol and
subsequently detached from the supporting silicon wavers by floating
off in hot water and transferred onto Cu grids~\cite{Murakami2006}.
The nematic order as well as optical
properties~\cite{Murakami2005a,Kramberger2007} and local morphology in
transmission electron microscopy of VA-SWNT~\cite{Einarsson2007} have
been studied earlier. The VA-SWNT are aligned within
$\unit{25}\degree$ and typically packed in small bundles with less
than 10 nanotubes, each with a diameter of about
$\unit{2}\nano\meter$. The angle resolved loss function of the VA-SWNT
was measured in a purpose built EELS spectrometer~\cite{Fink1989}.
Earlier comparative EELS studies were performed on a cleaved single
crystal~\cite{Marinopoulos2002} of graphite or bundled and
magnetically aligned SWNT~\cite{Liu2001}. In the present study we set
an energy and momentum resolution of $\unit{200}\milli\electronvolt$
and $\unit{0.05}\iA$ at a primary incidence energy of
$\unit{172}\kilo\electronvolt$. Concerning our \textit{ab initio}
simulations for isolated single- and double-layer graphene, we start
from DFT-LDA ground state calculations~\cite{CMS_Gonze_2002}, using a
plane-wave basis set~\footnote{We used 3481 k-points and an energy
  cutoff of 28 Hartree.} and norm--conserving pseudopotentials of
Troullier-Martins type \cite{Troullier1991}.  The loss function is
determined within the random phase approximation (RPA) using the
DP-code~\footnote{http://www.dp-code.org; V. Olevano, {\it et al.},
  unpublished.}. The local field effects (LFE) that originate from the
induced Hartree potential are taken into account, comprising in-plane
local fields with spatial variations on the atomic scale (Umklapp
effects).  Moreover, all contributing valence-electron bands ($\pi$
and $\sigma$ as well as empty states) are included.
\begin{figure}[htb]
\includegraphics[width=1\linewidth,clip]{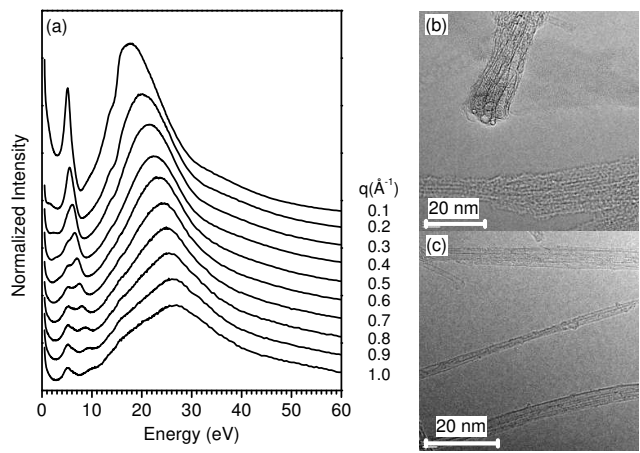}
\caption{ (a) measured loss function of freestanding VA-SWNT at
  equidistant $q$ from $\unit{0.1}\iA$ (top) to $\unit{1.0}\iA$
  (bottom). And TEM micrographs of the cross section (b) and side view
  (c) of the thin bundled VA-SWNT}
\end{figure}
Moving on to our EELS measurements on VA-SWNT, we first inspect
the loss function at the lowest momentum transfer
($\unit{0.1}\iA$), depicted topmost in Fig.~1a. We observe peaks
corresponding to the $\pi$ and the more structured $\pi+\sigma$
plasmon at $\unit{5.1}\electronvolt$ and
$\unit{17.6}\electronvolt$, respectively. These values are
remarkably low for sp$^2$ hybridized carbon
\cite{Liu2001,Knupfer1999} which indicates minimal macroscopic
screening and therefore fits well to the morphological
investigations of \citet{Einarsson2007}. TEM micrographs in
Fig.~1b\&c show the cross section and side view of this thin
bundled VA-SWNT. For increasing momentum transfer, the loss
spectra change significantly (shown up to $\unit{1.0}\iA$).
Apparently, both, the $\pi$ and $\pi\!+\!\sigma$ plasmon split
into two distinct contributions; one is localized like in a
molecule and another is dispersive like in a solid. We interpret
the localized response as a contribution with momenta
perpendicular to the axis, where an isolated quantum wire becomes
a confined dot. The dispersive response belongs to momenta
parallel to the nanotube axis, where we encounter an isolated one
dimensional wire.  The simultaneous, but distinct, observation of
these two intrinsic aspects of nanotubes - being {\it solid-like}
at the one hand and {\it molecule-like} at the other - is a
fingerprint of fully individualized viz. ideal wires in terms of
plasmon excitations. Following this interpretation, the
extrapolation of the corresponding $\pi$ plasmon position to the
optical limit ($q\!\to\!0$) predicts values of
$\unit{4.6}\electronvolt$ and $\unit{5.1}\electronvolt$ for the
on-axis and perpendicular component, respectively (see Fig.~2).
This is in excellent agreement with the optical absorption
findings of \citet{Murakami2005a}.
\begin{figure}[htb]
\includegraphics[width=1\linewidth]{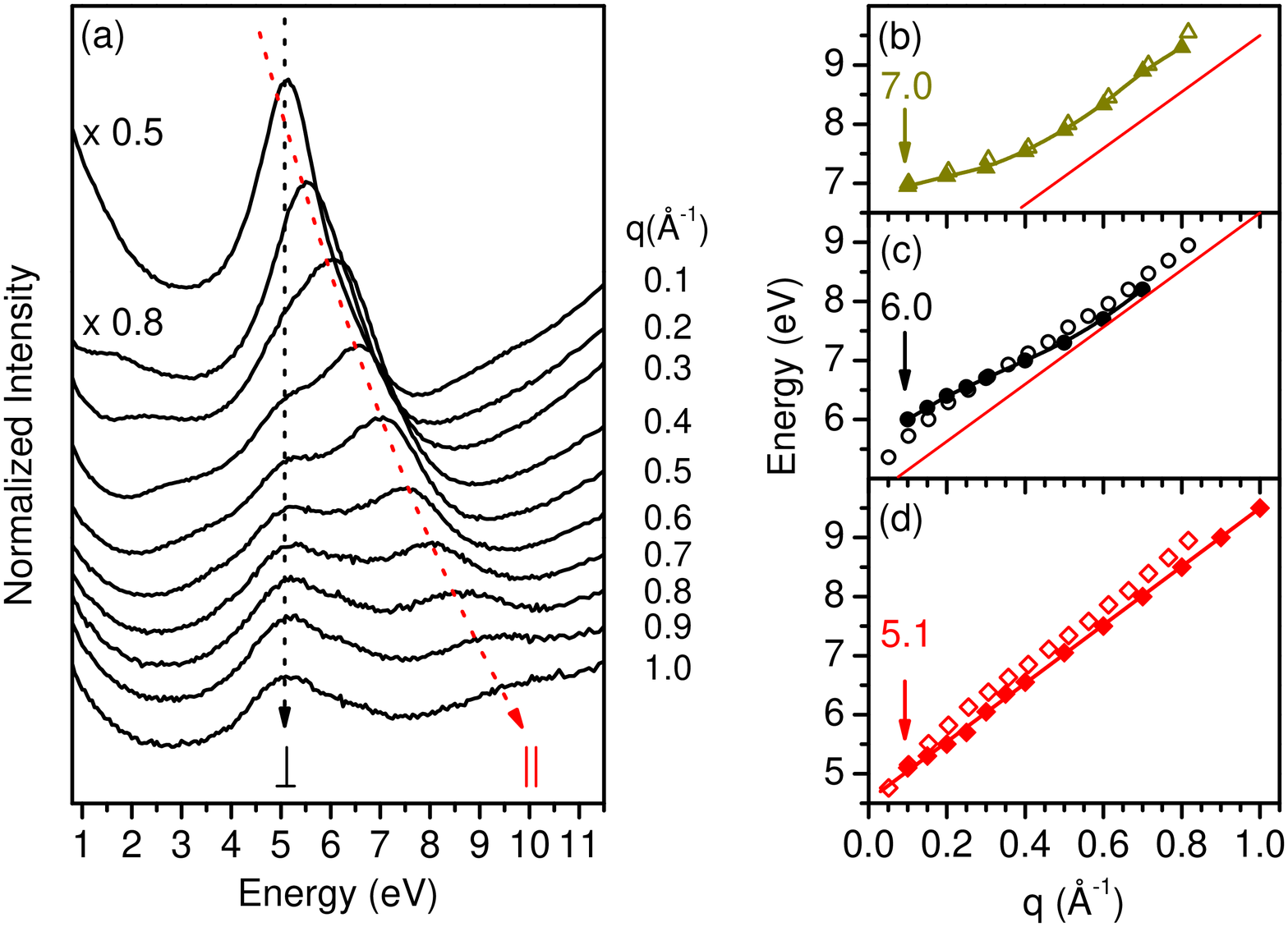}
\caption{ (color online) (a) loss function of the $\pi$ plasmon region
  at equidistant $q$ ranging from $\unit{0.1}\iA$ (top) to
  $\unit{1.0}\iA$ (bottom). Right stack: observed (filled symbols) vs.
  calculated (open symbols) $\pi$ plasmon dispersion for (b) graphite,
  (c) bundled SWNT \cite{Liu2001} vs. double layer graphene and (d)
  VA-SWNT vs. graphene. The calculations are averaged over the two
  in-plane directions $\Gamma$K and $\Gamma$M.  }
\end{figure}
Focusing on the dispersive on-axis part of the $\pi$ plasmon we
find a strikingly linear behavior (filled diamonds in Fig.~2d)
up to about $\unit{1}\iA$, covering more than one third of the
Brillouin zone. The linear behavior is totally different from the
parabolic $\pi$ plasmon dispersion of bulk graphite (filled
triangles in Fig.~2b), while bulk aligned SWNT (filled circles
\cite{Liu2001} in Fig.~2c) represent an intermediate situation.
Therefore, the plasmon dispersion at low $q$ is indeed a
fingerprint for the degree of isolation. The strictly linear
dispersion of the $\pi$ plasmon is only visible in isolated tubes.
\par
Although this dispersion might resemble the linear dispersion of
\lq Dirac electrons\rq\ in graphene, the structures seen in our
VA-SWNT measurements ($4$--$\unit{9}\electronvolt$) are clearly
outside the energy range of linear band dispersion in graphene.
Nevertheless, we will show in the following that graphene {\it is}
indeed {\it the} system to be used for the interpretation of the
on-axis $\pi$ plasmon dispersion of VA-SWNT. Our \textit{ab
initio} calculations allow us to isolate unambiguously the
features of this prototype system.
\par
To this aim, the loss function
$-\!\operatorname{Im}\epsilon^{-1}(\vec{q},\omega)$ of graphene
was calculated for different momentum transfers $\vec{q}$ along
the in-plane $\Gamma$M and $\Gamma$K directions, for values of
$q=|\vec{q}|$ ranging from $0.05$ to $\unit{0.8}\iA$. Starting
with the bare RPA (without LFE), the loss function is determined
by the independent-particle response function $\chi_0$, as in that
case
$\operatorname{Im}\epsilon^{-1}\propto\operatorname{Im}\chi_0$.
The resulting spectra can hence be interpreted as a sum of
independent transitions, which are directly related to the
bandstructure. Fig.~3a shows a typical spectrum for
$q=\unit{0.41}\iA$. In the low energy ($<\unit{10}\electronvolt$)
region, only transitions between the $\pi$ and $\pi^*$-band
contribute to the spectrum, which consists of three peaks in
$\Gamma$M direction (thin green line) but only two peaks for
$\Gamma$K (not shown).  In Fig.~3b the corresponding dispersions
are depicted (green solid and blue dotted lines, respectively).
The first peak arising from transitions within the \lq Dirac
cone\rq\ at K starts for the lowest $q$ at
$\unit{0.5}\electronvolt$ and disperses {\it linearly} to
$\unit{4.0}\electronvolt$ for the largest $q$.  The second peak,
only visible for $\Gamma$M is a weaker structure around
$\unit{4}\electronvolt$ which shows almost no dispersion.  The
last peak starting at $\unit{4.0}\electronvolt$ shows a quadratic
dispersion at small $q$. It can be attributed to transitions near
the edge of the Brillouin zone close to M. This peak is almost
undetectable in the joint density of states (see dotted curve in
Fig.~3a) but is strongly enhanced by matrix elements: already at
the independent particle level, bandstructure alone is not
sufficient to describe the spectrum completely.
\begin{figure}[htb]
\includegraphics[width=1\linewidth]{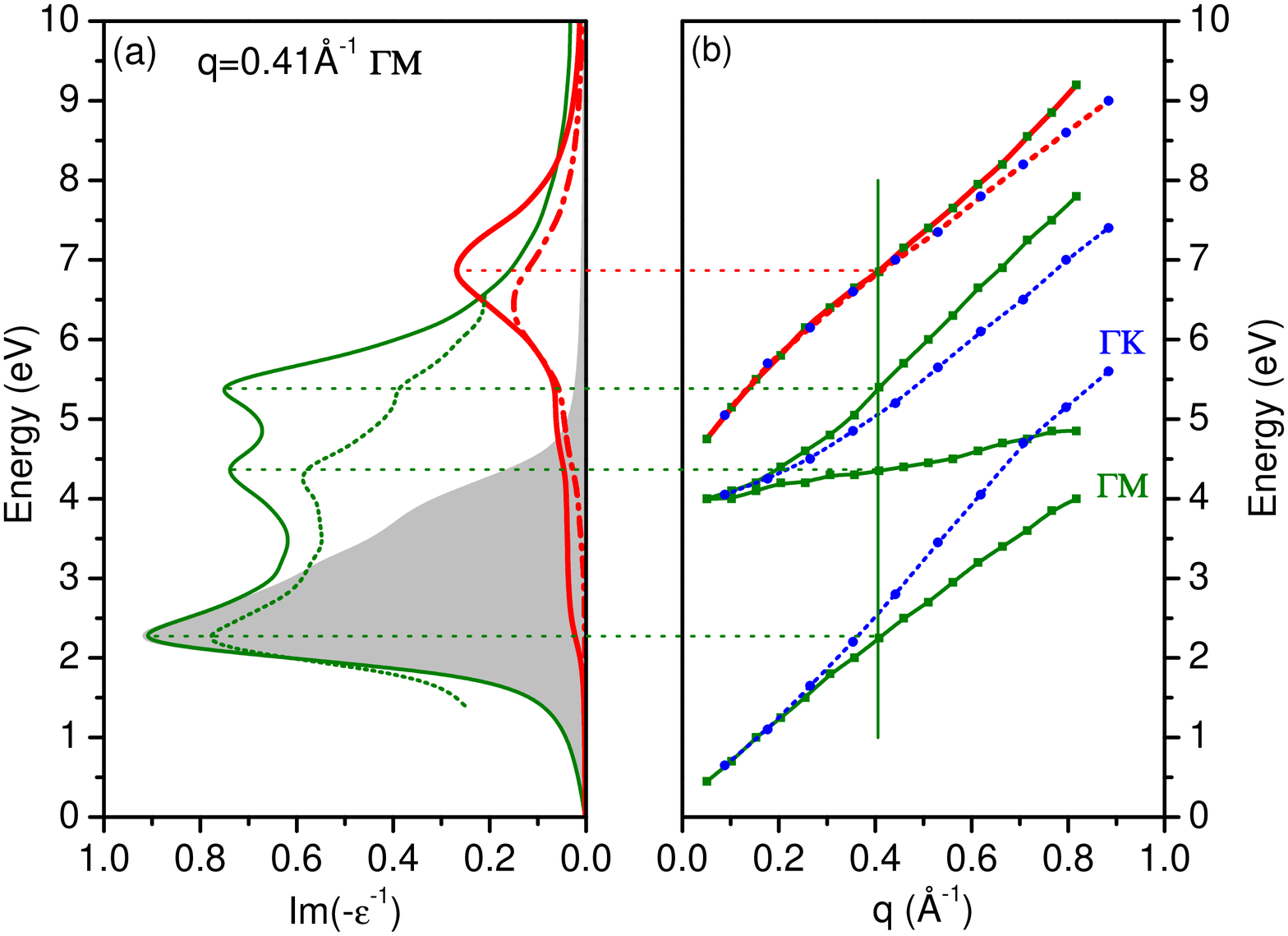}
\caption{(color online) (a) loss function of graphene at
  $q=\unit{0.41}\iA$ along $\Gamma$M calculated from the JDOS (green
  dots), within the bare RPA (green) and in RPA including LFE (red,
  thick). The latter changes significantly, when transitions next to
  the K point (shaded area), are excluded (red, dot-dashed). (b)
  dispersion of the peaks in the loss function for different momenta
  along $\Gamma$M (solid lines \& squares) and $\Gamma$K (dotted lines
  \& circles)}
\end{figure}
\par
When LFE are included in the calculation one determines
$\epsilon^{-1}=1+v\chi$ from the full response function $\chi =
\chi_0 + \chi_0 v \chi$, where the bare Coulomb interaction $v$
reflects the variation of the Hartree potential.  The inclusion of
this term accounts for LFE and changes the results drastically
(thick red lines): whereas induced microscopic components have
only little effect on the in-plane excitations in bulk
graphite~\cite{Marinopoulos2004}, LFE are of major importance for
the isolated sheets. Most importantly, they completely suppress
the linearly dispersing low energy structure as well as the very
weakly dispersing second peak. Instead, the peak starting at
$\unit{4}\electronvolt$ is blue shifted by about
$\unit{0.8}\electronvolt$ and becomes the dominant structure in
the spectrum. Its dispersion is strongly modified: LFE transform
the formerly quadratic dispersion into an almost linear one (red
line in Fig.~3b).  One can understand the LFE as a mixing of
transitions that occurs in the inversions when one solves the
screening equation for $\chi = (\chi_0^{-1} - v)^{-1}$. Therefore,
the resulting spectra should consist of mixtures of the formally
distinct peaks. This can involve a significant energy range. It is
therefore most interesting, to analyze whether the linearly
dispersing low energy peak has considerable influence on the
spectra including LFE. By choosing which transitions we include in
$\chi_0$, we compare the spectra with and without the
contributions from the the linear region of the $\pi$-bands around
the K point (i.\,e. low energy transitions). In the bare RPA loss
function $\chi_0$ this region gives rise to the shaded low energy
peak in Fig.~3a. Despite the very different energy ranges, the
final loss function after inclusion of LFE is indeed significantly
affected by the inclusion (red solid line) or exclusion (red
dot-dashed line) of these transitions: in the latter case the
dominant structure is strongly reduced (the integrated intensity
decreases by more than $30\%$) and is red shifted by about
$\unit{0.4}\electronvolt$. There are, hence, considerable
contributions from low energy transitions in the LFE corrected
plasmon response. With the mixing of transitions of different
energies also the different dispersion relations mix; therefore,
the resulting almost linear dispersion is indeed a superposition
of the dispersion of the main structures in the spectrum,
including that resulting from the \lq Dirac cone\rq.
\par
Figure~2d shows the resulting almost linearly dispersing $\pi$ plasmon of
graphene (open diamonds) together with the on-axis $\pi$ plasmon
of VA-SWNT (filled diamonds). The graphene plasmon reproduces
qualitatively, and even quantitatively, the experimental findings
on individualized VA-SWNT. Since our results are completely
parameter free, we can conclude that beyond qualitative arguments
concerning the tight relation of bandstructures and a similar
mechanism of the LFE, graphene can be studied in order to get
insight and quantitative information about VA-SWNT, and vice
versa.
\par
We now move on to elucidate the interacting case of bundled tubes
as shown in Fig.~2c. Experiments on bulk SWNT (filled circles
\cite{Liu2001}) revealed an asymptotic dispersion relation: the
$\pi$ plasmon is initially shifted to higher energies at small $q$
before it approaches the linear dispersion of VA-SWNT at large
$q\gtrsim\unit{0.5}\iA$. We employ bilayer graphene as an
appropriate model system for representing a typical next neighbor
situation. Indeed, the calculated $\pi$ plasmon dispersion of
isolated double layer graphene with an interlayer distance of
$d=\unit{3.3}\angstrom$ (as in graphite) shows the same overall
behavior (open circles) as the measurements on bundled tubes: we
find a transition from a regime, where the Coulomb interaction
$v\propto q^{-2}$ yields long range contributions involving
neighboring layers, to a situation where $q$ is sufficiently large
to confine main interactions to the same plane. In particular, we
studied at which interlayer distance $d$ the crossover from
interacting to non-interacting sheets occurs. For small
$q=\unit{0.1}\iA$, a distance of $\unit{30}\angstrom$ is necessary
in order to suppress the influence of neighboring sheets on the
spectra, while for $q=\unit{1.3}\iA$ the interlayer distance can
be reduced to $\unit{7}\angstrom$. In close analogy, the distinct
$\pi$ plasmon dispersion of bundled SWNT represents a smooth
dimensional crossover from three dimensional bundles to one
dimensional separated wires.  Hence high $q$ measurements are
applicable to probe the intrinsic properties of individual tubes
in bundles.
\par
Summarizing, we observe distinct $\pi$ plasmon dispersions in bulk
graphite, bundled SWNT and individualized VA-SWNT. Owing to the
individualization in the VA-SWNT we find a localized perpendicular and
a strictly linear on-axis $\pi$ plasmon dispersion. Our {\it ab
  initio} studies uncover drastic changes of the spectral RPA response
of graphene upon the inclusion of local field effects.  These LFE
account for a linearly dispersing $\pi$ plasmon in isolated graphene.
If a system can be considered to be isolated or not depends strongly
upon the momentum transfer $q$. In bundled SWNT a transition from an
interacting to a quasi non-interacting regime for large $q$ occurs and
leads to an asymptotic dispersion relation.  Measurements on VA-SWNT
assisted by calculations on graphene-based systems can hence discern
the contributions of the building blocks and their interaction, and
show that the study of a prototype system of this kind can be used to
obtain insight into the collective electronic excitations of related
materials.
\par
{\bf Acknowledgements:} This work was supported by the DFG PI 440
3/4, the EU's 6th Framework Programme through the NANOQUANTA
Network of Excellence (NMP4-CT-2004-500198) and by the ANR
(project NT0S-3 43900). Computer time was provided by IDRIS
(project 544). C.~K. acknowledges the \textit{IMPRS for Dynamical
Processes in Atoms, Molecules and Solids}. R.~H. thanks the
Dr.~Carl~Duisberg-Stiftung and C'Nano~IdF (IF07-800/R).
We thank S.~Leger, R.~H{\"u}bel, and R.~Sch{\"o}nfelder for technical assistance.

\end{document}